**Extremely high upper critical field in BiCh$_2$-based (Ch: S and Se) layered superconductor LaO$_{0.5}$F$_{0.5}$BiS$_{2-x}$Se$_x$ ($x$ = 0.22 and 0.69)**

Kazuhisa Hoshi[1], Ryosuke Kurihara[2], Yosuke Goto[1], Masashi Tokunaga[2], and Yoshikazu Mizuguchi[1]*

[1]Department of Physics, Tokyo Metropolitan University, 1-1 Minami-osawa, Hachioji, Tokyo 192-0397, Japan
[2]The Institute for Solid-State Physics, University of Tokyo, 5-1-5 Kashiwanoha, Kashiwa, Chiba 277-8581, Japan

Corresponding author: Y. Mizuguchi (mizugu@tmu.ac.jp)

**Abstract**

Centrosymmetric compounds with local inversion symmetry breaking have tremendously interesting and intriguing physical properties. In this study, we focus on a BiCh$_2$-based (Ch: S, Se) layered superconductor, as a system with local inversion asymmetry, because spin polarisation based on the Rashba–Dresselhaus-type spin-orbit coupling has been observed in centrosymmetric BiCh$_2$-based LaOBiS$_2$ systems, while the BiCh$_2$ layer lacks local inversion symmetry. Herein, we report the existence of extremely high in-plane upper critical fields in the BiCh$_2$-based system LaO$_{0.5}$F$_{0.5}$BiS$_{2-x}$Se$_x$ ($x$ = 0.22 and 0.69). The superconducting states are not completely suppressed by the applied magnetic fields with strengths up to 55 T. Thus, we consider that the in-plane upper critical field is enhanced by the local inversion symmetry breaking and its layered structure. Our study will open a new pathway for the discovery of superconductors that exhibit a high upper critical field by focusing on the local inversion symmetry breaking.

**Introduction**

Superconducting states are destroyed when a magnetic field is applied to the superconductor, and the maximum field is the upper critical field $B_{c2}$ in type-II superconductors. Superconductors that yield high $B_{c2}$ have attracted significant attention [1–8]. $B_{c2}$ is determined by two distinct pair-breaking effects: the paramagnetic pair-breaking effect and orbital pair-breaking effect. The paramagnetic and orbital pair-breaking effects arise from spin polarization which is attributed to the Zeeman effect and vortex formation, respectively. The lower one between the two limits, i.e. between the paramagnetic and orbital limit, is observed as $B_{c2}$ of the examined material. The former effect depends on the pairing symmetry of superconductivity. In spin-singlet superconductors, the limit by the paramagnetic pair-breaking effect is given by the relation $B_P = \Delta/\sqrt{g}\mu_B$; if we assume the weak-coupling limit $\Delta = 1.76 k_B T_c$ and $g = 2$ (where $\Delta$, $k_B$, $T_c$, $\mu_B$, and $g$ are the superconducting gap, the Boltzmann constant, the superconducting transition temperature, the Bohr magneton, and $g$-factor, respectively), the paramagnetic limit (Pauli limit) is given by the relation $B_P = 1.86 T_c$. In contrast, the paramagnetic pair-breaking effect is absent in a specific magnetic field in spin-triplet superconductors. Several uranium-based superconductors such as UCoGe, URhGe, and UTe$_2$ exhibit large $B_{c2}$ values, which remarkably exceed the Pauli limit expected from the aforementioned relation, owing to the spin-triplet superconductivity [1–3]. In non-centrosymmetric heavy fermion superconductors, the paramagnetic pair-breaking effect is suppressed by the unique spin texture of the Rashba-type spin-orbit coupling [4, 5]. The orbital limit is also enhanced in heavy-fermion superconductors because a large effective mass leads to a small coherence length. The orbital pair-breaking effect is also strongly suppressed in two-dimensional (2D) systems and layered superconductors in a magnetic field parallel to the in-plane direction. This can be understood by the out-of-plane coherence length $\xi_\perp$, which is smaller than the thickness in 2D superconductors, and the large out-of-plane effective mass, which originates from the anisotropic layered structure. In 2D superconductors of transition-metal dichalcogenides, wherein the orbital pair-breaking effect is almost quenched and the paramagnetic pair-breaking effect is predominant, the Zeeman-type spin-orbit coupling originating from the broken in-plane inversion symmetry plays a significant role in superconductivity [6–8]. The Zeeman-type spin-orbit coupling provides an intriguing Ising state which causes the enhancement of the $B_{c2}$.

Recently, theoretical studies have predicted that even if global inversion symmetry is preserved in a material, the breaking of local inversion symmetry leads to interesting physical phenomena such as parity-mixed superconductivity [9], stabilised odd-parity superconductivity [10], and the possibility of topological crystalline superconductivity with global centrosymmetric systems [11]. Rashba-type spin-orbit coupling induced by locally non-centrosymmetric systems is expected to suppress the paramagnetic pair-breaking effect and enhance the $B_{c2}$ [12]. An anomalous $B_{c2}$ was observed in artificial superlattices composed of heavy-fermion superconductor CeCoIn$_5$ and normal metal YbCoIn$_5$, wherein the inversion symmetry is locally broken at the interface between the CeCoIn$_5$ layer, despite the fact that the global inversion symmetry is present [13]. The Rashba-type spin-orbit coupling can be controlled by tuning the thickness modulation [14–16]. The MoS$_2$ bilayer (2H-MoS$_2$) also has global inversion symmetry, while the inversion symmetry within the individual layers is locally broken. The Zeeman-type spin-orbit interaction and Josephson coupling between the layers can be controlled by tuning the carrier concentration in the individual layers [17]. Moreover, the complex-strip phase, which has been predicted on the in-plane $B_{c2}$ in the multilayer systems with locally broken inversion symmetry, was confirmed in the bilayer and trilayer of a 2D superconductor [18]. The transition from the complex-strip phase to the helical phase with layer-dependent Rashba spin-orbit coupling was proposed in its systems [18]. More recently, a transition between two different superconducting phases and extremely high $B_{c2}$ has been observed in a heavy-fermion superconductor CeRh$_2$As$_2$, which possessed local inversion symmetry breaking at the cerium sites [19].

The typical BiCh$_2$-based (Ch: S and Se) system REOBiCh$_2$ (RE: rare earth), which is the target system of this study, has a layered crystal structure composed of REO blocking and BiCh$_2$ conducting layers [20, 21]. The parent phase of REOBiCh$_2$ is a band insulator. A partial F substitution at the O site generates electron carriers, and superconductivity emerges at low temperatures. The superconducting gap structure and pairing mechanism of the system have not been completely clarified. Several theoretical calculations suggest that anisotropic superconductivity, such as extended *s*-wave, *d*-wave, and *g*-wave states, are realised in BiCh$_2$-based superconductors [22–25]. However, early-stage experimental results have reported conventional fully gapped *s*-wave superconductivity; thermal conductivity, specific heat, and magnetic penetration depth measurements support fully gapped *s*-wave superconductivity [26–28]. In contrast, an anisotropic superconducting gap was observed in a laser angle-resolved photoemission spectroscopy (ARPES) apparatus [29]. An *s*-wave superconductor with accidental nodes was proposed based on a comparison with other experimental measurements, although the anisotropy is not contradictory to the *d*- and *g*-wave scenarios. The anisotropic *s*-wave gap structure observed by ARPES is almost consistent with the theoretical calculation, suggesting that the anisotropic gap arises from the Bi *p*-orbital degrees of freedom only with the use of purely attractive interactions [25]. Moreover, the absence of an isotope effect was observed, which also implies unconventional superconductivity [30]. The crystal structure of the BiCh$_2$-based superconductor possesses global inversion symmetry, whereas the inversion symmetry is locally broken in the BiCh$_2$ layer. Theoretical studies predict that hidden spin polarisation by local Rashba-type spin-orbit coupling should exist in the LaOBiS$_2$ system because of the site inversion asymmetry for Bi and S sites [31, 32]. The spin polarisation attributed to local Rashba spin-orbit coupling was observed by spin-ARPES (SARPES) for LaO$_{0.55}$F$_{0.45}$BiS$_2$ [33]. Furthermore, a high $B_{c2}$ was observed, which implies that the local inversion asymmetry can play a significant role in superconductivity [34]. However, research focused on local inversion asymmetry has not been extensively developed in BiCh$_2$-based superconductivity. In the BiCh$_2$-based superconductors, we selected LaO$_{0.5}$F$_{0.5}$BiS$_{2-x}$Se$_x$ ($x$ = 0.22 and 0.69) as the target materials to purely investigate the superconducting properties as the compounds have a bulk superconductivity nature [27] and do not contain $f$-electron elements in the REO blocking layer.

Herein, we show that extremely high in-plane $B_{c2}$ in the LaO$_{0.5}$F$_{0.5}$BiS$_{2-x}$Se$_x$ ($x$ = 0.22 and 0.69) where the inversion symmetry is locally broken in the BiCh$_2$ layer. The superconducting states were not completely suppressed by applied fields with strengths up to 55 T. The two pair-breaking effects, the paramagnetic pair-breaking effect and the orbital pair-breaking effect, are strongly suppressed by the local inversion symmetry breaking in the BiCh$_2$ layers and the layered structure.

**Results and discussion**

Figure 1a shows the crystal structure of the target system LaO$_{0.5}$F$_{0.5}$BiS$_{2-x}$Se$_x$ ($x$ = 0.22 and 0.69) which has a layered structure with tetragonal symmetry (*P*4/*nmm*). Although the crystal structure possesses global inversion symmetry (the symbol indicated with P in Fig. 1b shows the global inversion centre), the local inversion symmetry is broken in each BiCh$_2$ layer (dashed rectangles show each BiCh$_2$ layer in Fig. 1b). A partial Se substitution for the in-plane S site (Ch1 site) leads to the enhancement of the bulk nature of superconductivity and a specific heat jump is clearly observed [27]. Both the in-plane chemical pressure effect and carrier concentration have been revealed to be essential for the emergence of bulk superconductivity in the REOBiCh$_2$ system [21]. The $x$ value (Se concentration) was estimated using energy-dispersive X-ray (EDX) spectroscopy. The actual atomic ratio is almost consistent with the nominal value. Figures 1c and 1d show the temperature dependence of the in-plane resistivity $\rho_{ab}$ at a field strength of 0 T for $x$ = 0.22 and 0.69. The $T_c$ defined as the midpoint of the transition is 3.2 K and 4.1 K for $x$ = 0.22 and 0.69, respectively; the higher Se concentration causes the higher $T_c$. A weak upward behaviour was observed for $x$ = 0.22 and 0.69, which is similar to previous reports [34]. The amplitude of the upturn behaviour at low temperatures was suppressed by the Se substitution.

Figures 2a and 2b respectively show the temperature dependence of the in-plane resistivity $\rho_{ab}$ for $x$ = 0.22 at various fields parallel to the *ab*-plane and *c*-axis. The applied electric current was perpendicular to the magnetic field in $B \parallel c$ and parallel in $B \parallel ab$. In the case of $B \parallel c$, the superconducting states are immediately suppressed by the applied field. In contrast, in the case of $B \parallel ab$, the superconductivity is robust against the applied field. Figures 2c and 2d exhibit the field dependence of the $\rho_{ab}$ for $x$ = 0.22 from 3.0 K to 2.0 K with $B \parallel ab$ and $B \parallel c$. The $\rho_{ab}(B)$ data show a trend similar to that of $\rho_{ab}(T)$. Figures 2e-2h show the $x$ = 0.69 data. The behaviours of both $\rho_{ab}(T)$ and $\rho_{ab}(B)$ are similar to that of $x$ = 0.22. For $x$ = 0.69 with $B \parallel ab$, zero resistivity is retained up to 9 T for temperatures below 2.6 K. The temperature and field dependencies of $\rho_{ab}$ show a large anisotropic factor $\gamma$ for both $x$ = 0.22 and 0.69, thus indicating that the superconductors have strong anisotropic (two-dimensional) superconducting characteristics. $B_{c2}$ was estimated from the midpoint of the resistive transition of the $\rho_{ab}(B)$ data (see Supplementary Fig. 2); a similar criterion has been used in several studies [6, 7, 13, 14, 34]. Moreover, we plotted the $B_{c2}$ defined as the beginning of the resistive increase from zero resistivity of the $\rho_{ab}(B)$ data and from the $\rho_{ab}(T)$ data, respectively, in Supplementary Figs. 4 and 5.

To investigate the in-plane $B_{c2}$ variation at higher fields, we measured the field dependence of $\rho_{ab}$ for $x$ = 0.22 and 0.69 based on the use of pulsed high fields up to 55 T. Figures 3a and 3b show the $\rho_{ab}(B)$ data acquired by the pulsed fields parallel to the $ab$ plane from 4.2 K to ~0.47 K for $x$ = 0.22 and 0.69, respectively. Notably, the superconducting states are not completely destroyed up to 55 T at ~0.47 K for both $x$ = 0.22 and 0.69. Moreover, for $x$ = 0.69, the superconducting states survive at the highest field of 55 T even at 1.39 K. The gradient of the $\rho_{ab}(B)$ curves from zero resistivity to the normal state became smaller in the low-temperature and high-field regions for both samples. We defined the normal state resistivity $\rho_n$ as the black dashed lines in Figs. 3a and 3b because it is not largely changed by the scanning field. Thus, the strength of $B_{c2}$ was estimated from the midpoint of zero resistivity and $\rho_n$ (see Supplementary Fig. 2). We also plotted the $B_{c2}$ defined as the beginning of the resistive increase from zero resistivity of the $\rho_{ab}(B)$ curves by the pulsed field (see Supplementary Figs. 3 and 4). We summarise the $B_{c2}$ obtained by the static fields (Fig. 2) and pulsed fields (Fig. 3) for $x$ = 0.22 and 0.69 in Figs. 4a and 4b, respectively. The pulsed-field data are not contradictory to the static-field data because we used the same single crystal samples. Note that there is uncertainty in the field directions in the pulsed-field measurements to deviate slightly from the $ab$ plane because we could not use a rotator system in the setup; therefore, we added the error bar to the blue circles in Fig. 4. The upturn of $B_{c2}^{\parallel}$ at the low-temperature region indicates that $B_{c2}^{\parallel}(0)$ is higher than our minimum temperature value. The orbital limit $B_{\text{orb}}(0)$ is estimated to be 9.1 and 0.33 T within the in-plane and out-of-plane direction for $x$ = 0.22 and 17.2, and 0.58 T within the in-plane and out-of-plane direction for $x$ = 0.69 from the initial slope of $B_{c2}$ at $T_c$ based on the relation $B_{\text{orb}}(0) = 0.69T_c(-dB_{c2}/dT)_{T_c}$ in the dirty limit [36]. We describe the Werthamer–Helfand–Hohenberg (WHH) curves (dashed lines in Figs. 4a and 4b). The observed $B_{c2}^{\parallel}$ clearly exceeds the in-plane orbital limit $B_{\text{orb}}^{\parallel}(0)$ and the WHH curves are not suitable for $B_{c2}^{\parallel}$ at $x$ = 0.22 and 0.69. Moreover, the out-of-plane orbital limit $B_{c2}^{\perp}$ also deviates from the WHH curves (see the inset of Fig. 4), which is consistent with a previous report [27, 34]. For another pair-breaking effect, the Pauli limit is estimated from the relation $B_P = 1.86T_c$. Pauli limit is estimated to be 5.9 and 7.6 T for $x$ = 0.22 and 0.69, respectively; these values (black diamonds in Fig. 4) are much lower than the observed $B_{c2}^{\parallel}$. We performed linear fitting (solid lines in the inset of Fig. 4) to evaluate the $B_{c2}^{\perp}(0)$ because the WHH curves are not applicable to the $B_{c2}^{\perp}(T)$ for both samples. The $B_{c2}^{\perp}(0)$ values were estimated to be 0.57 T and 0.98 T for both $x$ = 0.22 and 0.69, respectively. The in-plane coherence length $\xi_{\parallel}$ was evaluated from the relation $B_{c2}^{\perp} = \Phi_0/2\pi\xi_{\parallel}^2$ based on $B_{c2}^{\perp}(0)$ obtained from the linear fitting. This led to $\xi_\parallel \sim 24$ and $\sim 18$ nm for $x = 0.22$ and 0.69, respectively. These values are comparable to those reported previously for $x = 0$ [27, 34]. To estimate the out-of-plane coherence length $\xi_\perp$, we used the $B^*$ defined as the intersection of the extrapolation of the $\rho_{ab}(B)$ curves at the minimum temperature of $\sim 0.47$ K and $\rho_n$ (see the insets of Fig. 3). Specifically, $B^*$ is expected to be the field which completely destroys the superconducting states at $\sim 0.47$ K. By using the values of $\xi_\parallel$ and $B^*$ at the minimum temperatures, we evaluated the out-of-plane coherence length $\xi_\perp$ at $\sim 0.47$ K from the relation $B_{c2}^\parallel = \Phi_0/2\pi\xi_\parallel\xi_\perp$. This resulted in $\xi_\perp \sim 0.22$ and $\sim 0.23$ nm for $x = 0.22$ and 0.69, respectively.

The angular ($\theta$) dependence of the $B_{c2}$ at 2.5 K for $x = 0.22$ and 3.5 K for $x = 0.69$ are displayed in Figs. 5a and 5b ($\theta$ represents the angle between the $c$-axis and the direction of the applied magnetic field). $B_{c2}(\theta)$ was also estimated from the middle point of the resistive transition. Generally, $B_{c2}(\theta)$ for layered superconductors is described by the anisotropic three-dimensional (3D) Ginzburg-Landau (GL) model or 2D Tinkham's formula. The anisotropic 3D GL model is given by the relation $(B_{c2}(\theta)\cos\theta/B_{c2}^\perp)^2 + (B_{c2}(\theta)\sin\theta/B_{c2}^\parallel)^2 = 1$ and explains $B_{c2}(\theta)$ anisotropic 3D systems. In contrast, the 2D Tinkham's formula is given by the $|B_{c2}(\theta)\cos\theta/B_{c2}^\perp| + (B_{c2}(\theta)\sin\theta/B_{c2}^\parallel)^2 = 1$ and describes $B_{c2}(\theta)$ in the case of the 2D systems [6]. The 2D Tinkham's formula exhibits cusp-like behaviour around the magnetic field parallel to the in-plane. The observed $B_{c2}(\theta)$ is not contradictory to the anisotropic 3D GL model (solid line in Fig. 5). The broadening of $B_{c2}(\theta)$ at $x = 0.69$ can be attributed to the flux flow and/or tiny single crystals with an inclination of a few degrees. The 2D Tinkham's formula (dashed line in Fig. 5) is not favourable for the broad $B_{c2}(\theta)$ behaviour. Furthermore, we performed the $B_{c2}(\theta)$ measurements at different temperatures (Supplementary Fig. 1), which is consistent with Fig. 5.

The paramagnetic pair-breaking effect and the orbital pair-breaking effect should be suppressed because the observed $B_{c2}^\parallel$ clearly exceeds both the Pauli and the orbital limits. For the paramagnetic pair-breaking effect, the strong coupling nature leads to the enhancement of the Pauli limit. While we assumed the validity of the weak-coupling limit in the above explanation, some experimental results in BiCh$_2$-based superconductors indicate that the strong coupling limit applies [27, 28]. However, even if we use the reported value of $\Delta = 2.25k_BT_c$ for the single crystal of LaO$_{0.5}$F$_{0.5}$BiSSe ($x = 1.0$) [27], the Pauli limit is much smaller than the observed $B_{c2}^\parallel$. A previous study proposed that the behaviour of in-plane $B_{c2}$ for LaO$_{0.5}$F$_{0.5}$BiS$_2$ ($x = 0$) can be explained by the two-gap nature [34]. By contrast, specific heat measurements suggested a single gap for the LaO$_{0.5}$F$_{0.5}$BiSSe ($x = 1.0$) single-crystal sample [27]. We consider that the single-gap scenario is likely valid for the present samples because the superconducting properties, including the bulk nature of superconductivity of our single crystals, are similar to those for $x = 1.0$. Therefore, we suggest that the breaking of the local inversion symmetry in the BiCh$_2$ layer leads to a large $B_{c2}^\parallel$. The Rashba-type spin-orbit coupling due to the lack of local inversion symmetry enhances the Pauli limit because the spin direction is locked onto the $ab$ plane and the spin texture protects the Cooper pairs from de-pairing against the applied fields [12, 14, 34, 37]. The local Rashba-type spin texture near the Fermi energy was observed by SARPES for LaO$_{0.55}$F$_{0.45}$BiS$_2$ [33]. Moreover, the spin-singlet and spin-triplet states can be mixed in a material by breaking the local inversion symmetry [9]. If the spin-triplet component is the predominant component in the superconductivity in the present phases, the paramagnetic pair-breaking effect may be absent in this system. In any case, the paramagnetic pair-breaking effect was largely suppressed by the local inversion symmetry breaking in the BiCh$_2$ layer.

The orbital limit should be enhanced by a layered structure. The Rashba-type spin-orbit coupling generally weakens the interlayer coupling and enhances the 2D nature of the superconductivity [12]. This situation allowed us to establish the Josephson vortex state in $B \parallel ab$. In these conditions, we can deduce that the Josephson vortices penetrate in the LaO blocking layer when the out-of-plane coherence length $\xi_\perp$ is smaller than the thickness of the blocking layer, in which the vortices may induce the orbital pair-breaking effect. However, $B_{c2}(\theta)$ in Fig. 5 indicates that the 2D superconductivity is not very strong because the anisotropic 3D GL model is well fitted to the $B_{c2}(\theta)$ data. Therefore, it is reasonable to expect that the crossover of the conventional Abrikosov and Josephson vortex states was realised in this system. This description is almost consistent with the upturn behaviour of $B_{c2}(T)$ at $T_c$ [17]. To confirm whether the angular-dependent $B_{c2}$ is changed by temperature, we may need to investigate $B_{c2}(\theta)$ at the lower-temperature region by using dilution systems in future work. Evaluation of the out-of-plane coherence length $\xi_\perp \sim 0.22$ ($x = 0.22$) and $\sim 0.23$ nm ($x = 0.69$) at the minimum temperature indicated that the outcomes are comparable to the LaO blocking layer thicknesses $\sim 0.26$ nm in the cases of both samples. The blocking layer thicknesses were estimated from powder X-ray diffraction at 298 K. We expect that $B_{c2}$ could be determined by the orbital pair-breaking effect from the Josephson vortices if we could confirm the change from the anisotropic 3D GL model to 2D Tinkham's formula in the lower temperature region.

Herein, we discuss the intriguing complex-strip phase, which is a unique property of multilayer systems with local inversion symmetry breaking [18, 38]. This phase is induced by the vortices penetrating the blocking layer and is thus regarded as the Josephson vortex state [18]. Although our system behaviours are not expected to be similar to the simple Josephson vortex states as discussed above and the upturn of $B_{c2}(T)$ at $T_c$ does not seem suitable for the complex-strip phase, the behaviours of the $B_{c2}(T)$ obtained by the pulsed field (high-field regions) are similar to the transition of the complex-stripe phase to the helical phase [18]. If the mixture of spin-singlet and spin-triplet states is significant in our system, it may be difficult for the complex-stripe phase to emerge in this system because the complex-stripe phase is in spin-singlet superconductivity without a mixture of spin-triplet superconductivity [38].

Finally, we briefly discuss the pairing symmetry of the superconductivity of the BiCh$_2$-based superconductor. Many experiments are consistent with fully gapped $s$-wave superconductivity [26–28], while the ARPES study found an anisotropic superconducting gap [29]. The breaking of local inversion symmetry may be a key factor in the solution of the puzzle because spin-singlet and spin-triplet superconductivity can be mixed in the local inversion symmetry breaking. We expect that additional work on the local inversion asymmetry in BiCh$_2$-based superconductors will reveal the pairing symmetry of superconductivity.

In conclusion, we have observed extremely high in-plane $B_{c2}$ for the centrosymmetric superconductor LaO$_{0.5}$F$_{0.5}$BiS$_{2-x}$Se$_x$ with local inversion symmetry breaking in the BiCh$_2$ layer. The superconducting states were not completely destroyed, even at the field strength of 55 T. The paramagnetic pair-breaking effect should be suppressed by the local Rashba-type spin-orbit coupling which arises from the lack of local inversion symmetry, and was observed by SARPES [33]. The orbital limit should be enhanced by the layered structure and strong local Rashba spin-orbit coupling. Local inversion symmetry breaking may be a clue for the solution of the pairing symmetry of superconductivity in the BiCh$_2$-based superconductor family. Our results pave the way for the exploration of superconductors which have a high $B_{c2}$, and also lead to the in-depth understanding of the relationship between superconductivity and local inversion symmetry breaking.

## Methods
### Single crystal growth

Single crystal samples of LaO$_{0.5}$F$_{0.5}$BiS$_{2-x}$Se$_x$ ($x$ = 0.25 and 0.75) were grown by a high-temperature flux method in a quartz tube in vacuum [39]. First, polycrystalline samples were prepared by a conventional solid-state reaction method to obtain a nominal composition. The obtained polycrystalline (0.4 g) samples were mixed with CsCl/KCl flux (5.0 g) at a molar ratio of 5:3, ground, and then sealed in a quartz tube in vacuum. The quartz tube was heated at 1223 K for 10 h, cooled to 873 K at a rate of -2 K /h, and furnace-cooled to 298 K. The quartz tube was opened in air, and the flux was dissolved in a quartz tube using pure water. The real atomic ratio of the single-crystal samples for the resistivity measurement was estimated by EDX analysis, and its value was consistent with the nominal composition. The sizes of the plate-like single crystals for resistivity measurement are 1.53 mm × 2.05 mm × ~0.03 mm and 0.6 mm × 0.89 mm × ~0.04 mm for $x$ = 0.22 and 0.69, respectively.

### Transport measurements

Resistivity measurements for both static and pulsed fields were performed using the conventional four-probe method. The resistivity measurements in a static field up to 9 T and 2 K were performed using a physical property measurement system (PPMS) with a horizontal rotator probe. Pulsed magnetic field measurements were performed at the Institute for Solid State Physics (ISSP) at the University of Tokyo.

**Acknowledgements**

We thank Y. Yanase for fruitful discussions. K.H. was supported by the Japanese Society for the Promotion of Science (JSPS) through a Research Fellowship for Young Scientists. This work was partly supported by KAKENHI (Nos. JP20J21627, 18KK0076), and the Tokyo Metropolitan Government Advanced Research (H31-1). The experiments under pulsed magnetic fields were performed under a joint research program of ISSP, Tokyo University, under a proposal of No. 202105-HMBXX-0055). We would like to thank Editage (www.editage.com) for English language editing.

**Author contributions**

K.H., Y.G. and Y.M. conceived the study. K.H. synthesised and characterised the single crystals. K.H. performed transport measurements using the PPMS and analysed the data. K.H., R.K., and M.T. performed the high-field measurements in the ISSP. All the authors contributed to the physical discussions. K.H. and Y.M. wrote the manuscript.

**Competing interests**

The authors declare no competing financial or non-financial interests.

**Data availability**

The data that support the findings of this study are available from the corresponding authors upon reasonable request.

Figures

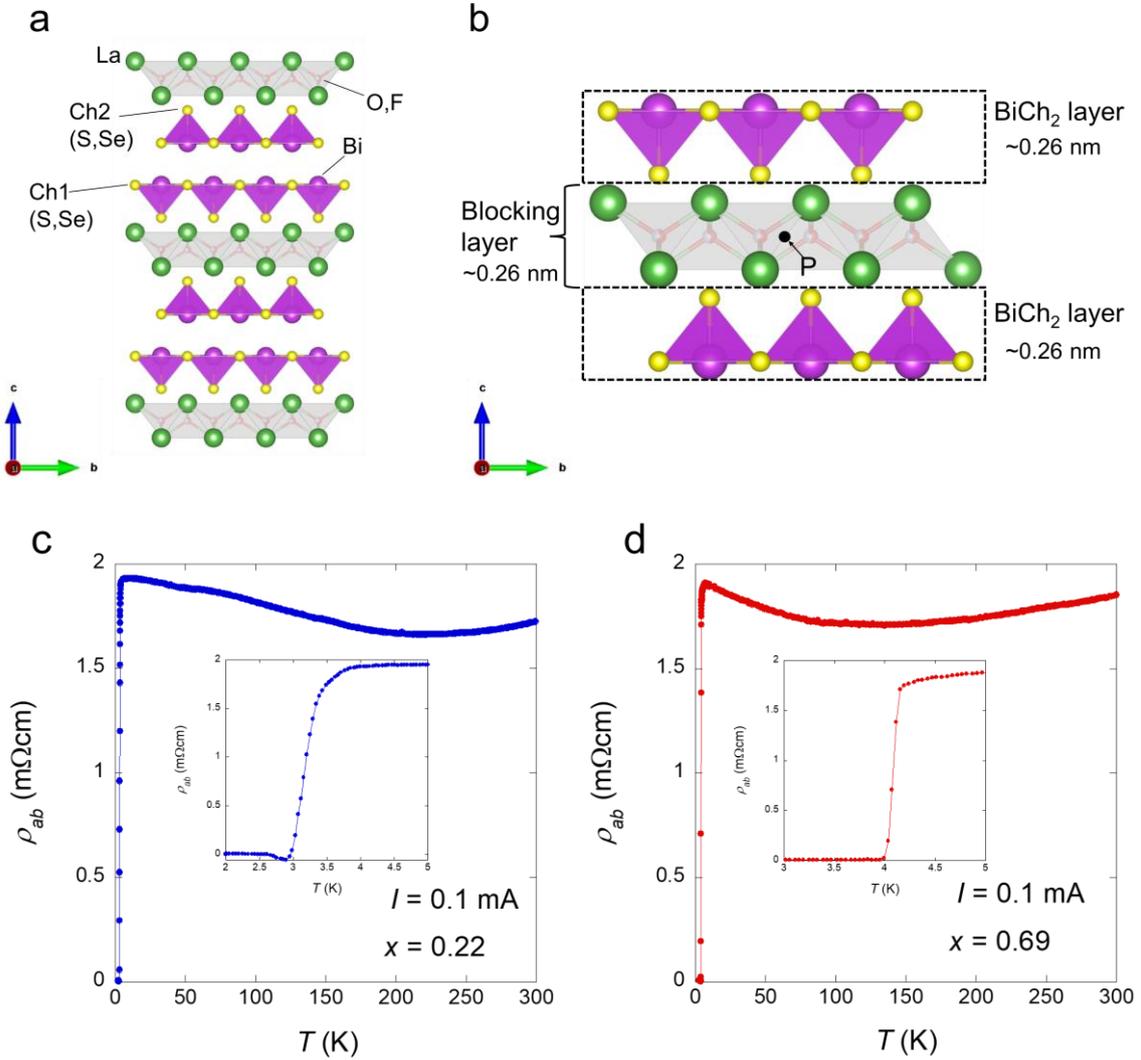

**Fig. 1** Crystal structure of $LaO_{0.5}F_{0.5}BiS_{2-x}Se_x$ and electrical resistivity in the absence of a magnetic field for $LaO_{0.5}F_{0.5}BiS_{2-x}Se_x$. **a** Crystal structure of $LaO_{0.5}F_{0.5}BiS_{2-x}Se_x$ with a space group of *P*4/*nmm* (No. 129). The crystal structure image is prepared by VESTA [35]. **B** Two $BiCh_2$ layers and LaO blocking layer. The inversion symmetry is locally broken in the each $BiCh_2$ layer. The symbol P in the LaO blocking layer denotes the global inversion centre for the $LaO_{0.5}F_{0.5}BiS_{2-x}Se_x$ system. The thicknesses of the $BiCh_2$ and the blocking layers were evaluated as the length between the centres of the two atomics by structural analysis of powder X-ray diffraction. **c, d** Temperature dependence of the resistivity in the absence of a magnetic field for $x = 0.22$ (**c**) and 0.69 (**d**). The insets show the enlarged resistivity curves near the superconducting transition. The superconducting transition temperature, $T_c$ is defined as the midpoint of the resistive transition and was observed at $T_c = 3.2$ K and 4.1 K.

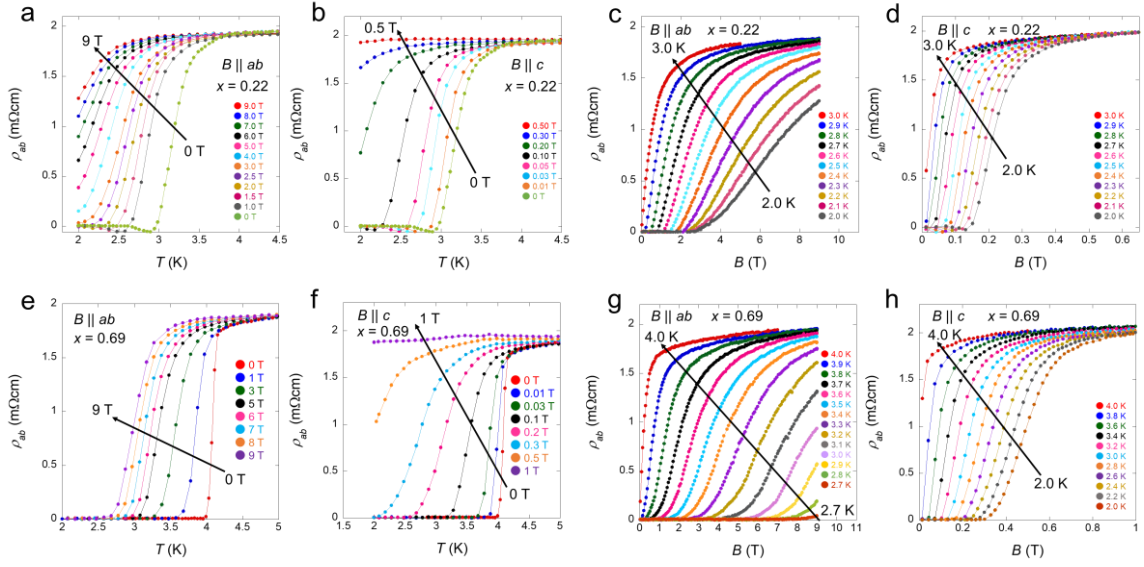

**Fig. 2** In-plane resistivity $\rho_{ab}$ for $x = 0.22$ and $0.69$ in the presence of a magnetic field. **a, b** Temperature dependence of the in-plane resistivity $\rho_{ab}$ in the presence of a magnetic field for $x = 0.22$ in $B \parallel ab$ (**a**) and $B \parallel c$ (**b**). The arrows denote the direction from low to high fields. **c, d** Field dependence of the $\rho_{ab}$ at different temperatures for $x = 0.22$ in $B \parallel ab$ (**c**) and $B \parallel c$ (**d**). The arrows denote the direction from low to high temperatures. **e–h** $\rho_{ab}(T)$ in $B \parallel ab$ (**e**) and $B \parallel c$ (**f**), and $\rho_{ab}(B)$ in $B \parallel ab$ (**g**) and $B \parallel c$ (**h**) for $x = 0.69$.

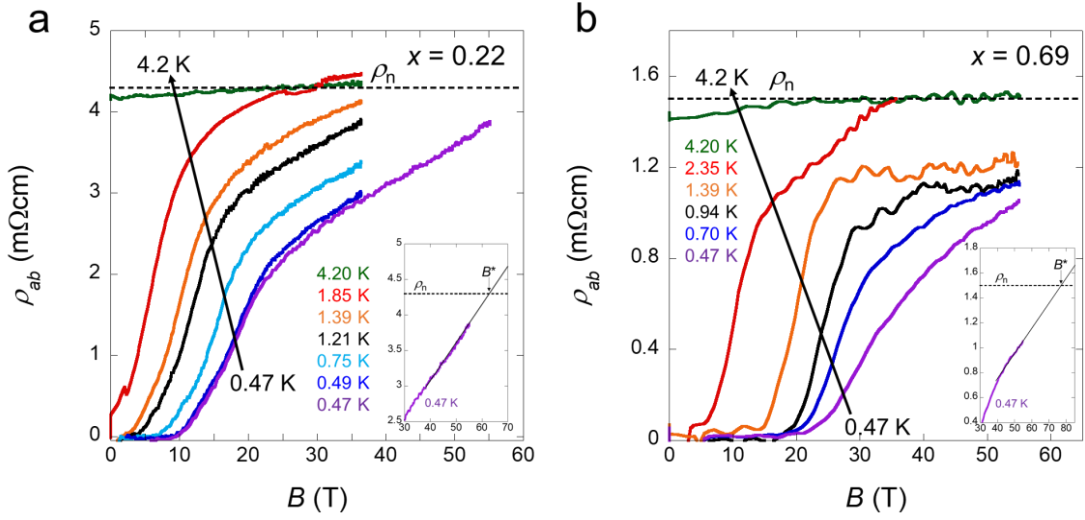

**Fig. 3** Field dependence of the resistivity by a pulsed magnetic field. **a, b** Field dependence of the in-plane resistivity in $B \parallel ab$ for $x = 0.22$ (**a**) and 0.69 (**b**). The $\rho_n$ (black dashed lines) denotes normal resistivity in the pulsed-field measurement. The insets show magnified views of the $\rho_{ab}(B)$ curves at ~0.47 K near $\rho_n$. $B^*$ is defined as the intersection of the extrapolation of the $\rho_{ab}(B)$ curves and the $\rho_n$.

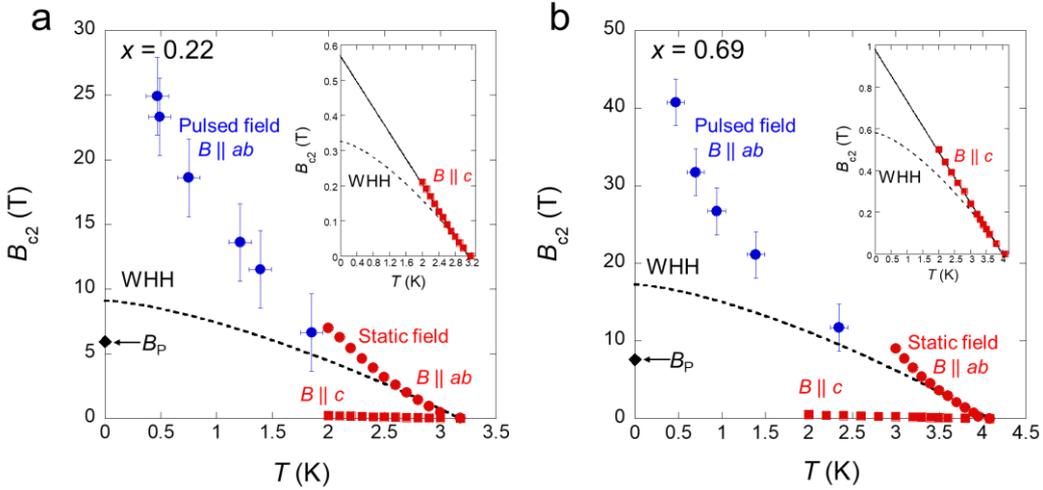

**Fig. 4** Huge upper critical field for locally non-centrosymmetric superconductor LaO$_{0.5}$F$_{0.5}$BiS$_{2-x}$Se$_x$ ($x = 0.22$ and 0.69). **a, b** Upper critical field as a function of temperature for $x = 0.22$ (**a**) and 0.69 (**b**). Red and blue circles show the $B_{c2}^{\parallel}(T)$ by static field and pulsed field, respectively. We added the error bars to the blue circles as it is possible for the applied field to slightly deviate from the $ab$ plane owing to the lack of rotator measurements. Additionally, there is uncertainty related to the stability of temperature. Red squares show the $B_{c2}^{\perp}(T)$ by the static field. The black dashed curves show the Werthamer–Helfand–Hohenberg (WHH) fits. Black diamonds denote the Pauli limit. The insets show magnified views of the low-field region for $B_{c2}^{\perp}(T)$. Black solid lines show linear fits.

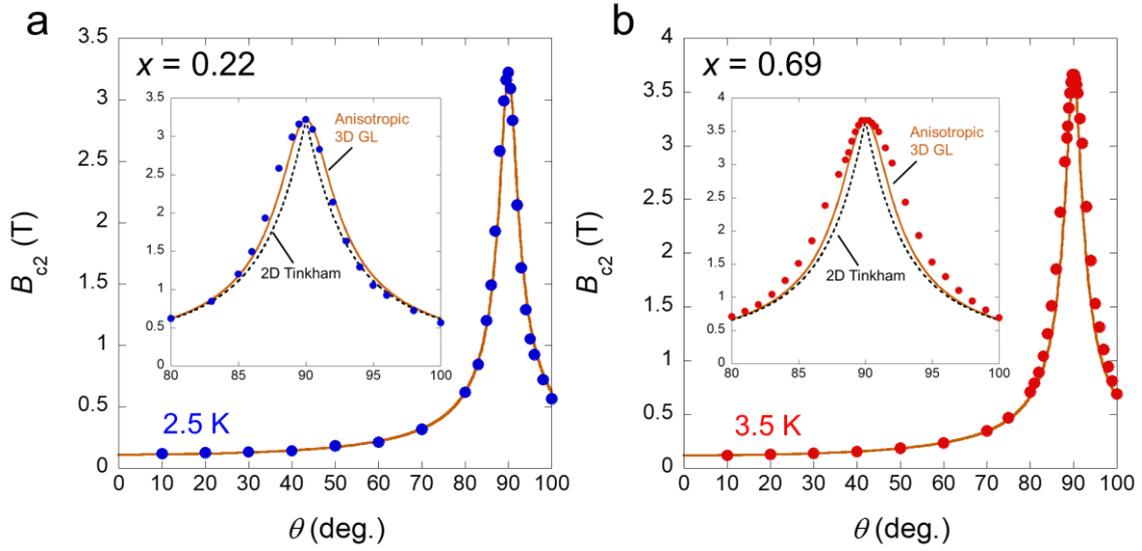

**Fig. 5** Three-dimensional (3D) nature of the upper critical field. **a, b** Angular $\theta$ dependence of the upper critical field at 2.5 K for $x = 0.22$ (**a**) and at 3.5 K for $x = 0.69$ (**b**). $\theta$ represents the angle between the $c$-axis and the direction of the applied magnetic field. The insets show magnified views of the region around $\theta = 90°$. The solid curves denote the anisotropic 3D Ginzburg-Landau (GL) model. The dashed curves show the two-dimensional (2D) Tinkham's formula.

**Extremely high upper critical field in BiCh$_2$-based (Ch: S and Se) layered superconductor LaO$_{0.5}$F$_{0.5}$BiS$_{2-x}$Se$_x$ ($x$ = 0.22 and 0.69)**

Kazuhisa Hoshi[1], Ryosuke Kurihara[2], Yosuke Goto[1], Masashi Tokunaga[2], and Yoshikazu Mizuguchi[1]

[1]Department of Physics, Tokyo Metropolitan University, 1-1 Minami-osawa, Hachioji, Tokyo 192-0397, Japan
[2]The Institute for Solid-State Physics, University of Tokyo, 5-1-5 Kashiwanoha, Kashiwa, Chiba 277-8581, Japan

### S.1 Angular dependence of $B_{c2}$ at different temperatures from the main text

We investigated the angular dependence of the upper critical fields for $x$ = 0.22 at 2.3 K and $x$ = 0.69 at 3.3 K, respectively. The anisotropic 3D GL model is relatively fitted for the angular-dependent $B_{c2}$, which is consistent with Fig. 5 in the main text.

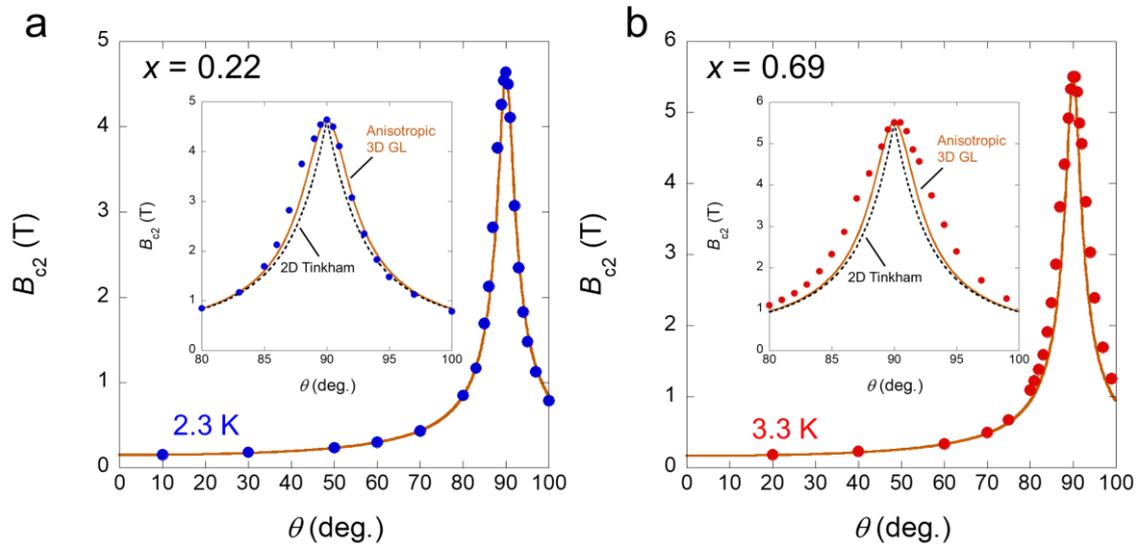

**Supplementary Fig. 1** Three-dimensional (3D) nature of the upper critical field. **a, b** Angular $\theta$ dependence of the upper critical field at 2.3 K for $x$ = 0.22 (**a**) and at 3.3 K for $x$ = 0.69 (**b**). $\theta$ represents the angle between the $c$-axis and the direction of the applied magnetic field. The insets show magnified views of the region around $\theta$ = 90°. The solid curves denote the anisotropic 3D Ginzburg-Landau (GL) model. The dashed curves show the two-dimensional (2D) Tinkham's formula.

**S.2 Estimation of $B_{c2}$**

In this work, we estimated the upper critical fields as the midpoint of the resistive transition of the $\rho_{ab}(B)$ curves. Many studies have approved of this criterion as the upper critical field [1-6]. We show the criteria to estimate the upper critical fields in Supplementary Fig. 2. Moreover, we estimated upper critical fields from the temperature where resistivity begins to increase from zero resistivity. We show examples of the criteria as the beginning of the resistive increase from zero resistivity of the $\rho_{ab}(B)$ curves in Supplementary Fig. 3. The upper critical fields estimated from the Supplementary Fig. 3 are displayed in Supplementary Fig. 4. The observed upper critical fields for both $x = 0.22$ and $x = 0.69$ by the pulsed fields clearly exceed Pauli limits and deviate from WHH curves although the upper critical fields for $x = 0.22$ by static field (red closed circles in Supplementary Fig. 4a) are almost consistent with the WHH curve. The upper critical fields by static fields for $x = 0.69$ (red closed circles in Supplementary Fig. 4b) show upward behaviour, which is similar to Fig.4b in the main text.

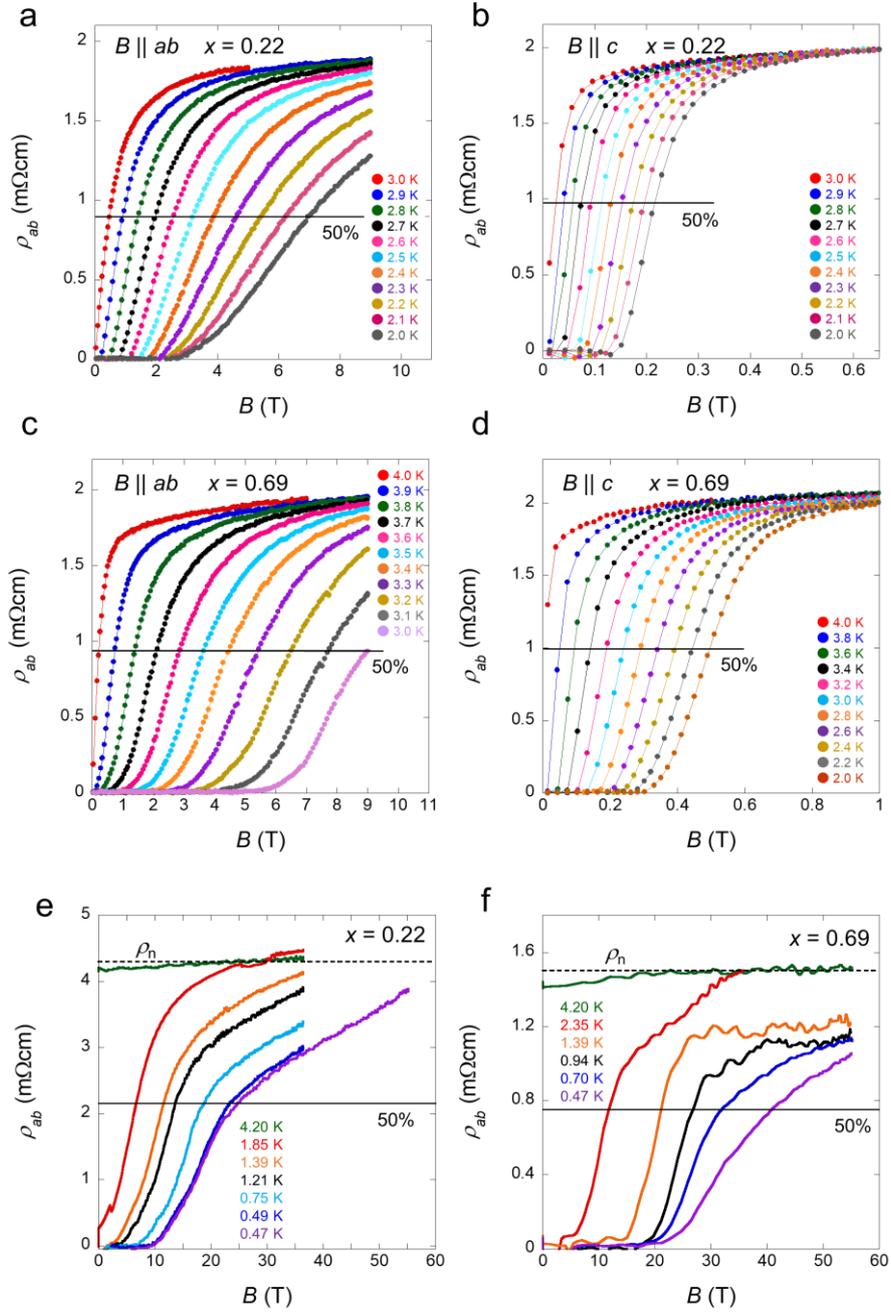

Supplementary Fig. 2 Estimation of $B_{c2}$ form the midpoint of the resistive transition of the $\rho_{ab}(B)$ curves. **a–f** Field dependence of the resistivity by static field for $x = 0.22$ in the fields parallel to the *ab*-plane (**a**) and *c*-axis (**b**), for $x = 0.69$ in the fields parallel to the *ab*-plane (**c**) and *c*-axis (**d**), and by pulsed field for $x = 0.22$ (**e**) and $x = 0.69$ (**f**). The black solid lines of 50% indicate the midpoint of the resistive transition of the $\rho(B)$ curves to estimate the upper critical fields for Fig. 4 in the main text.

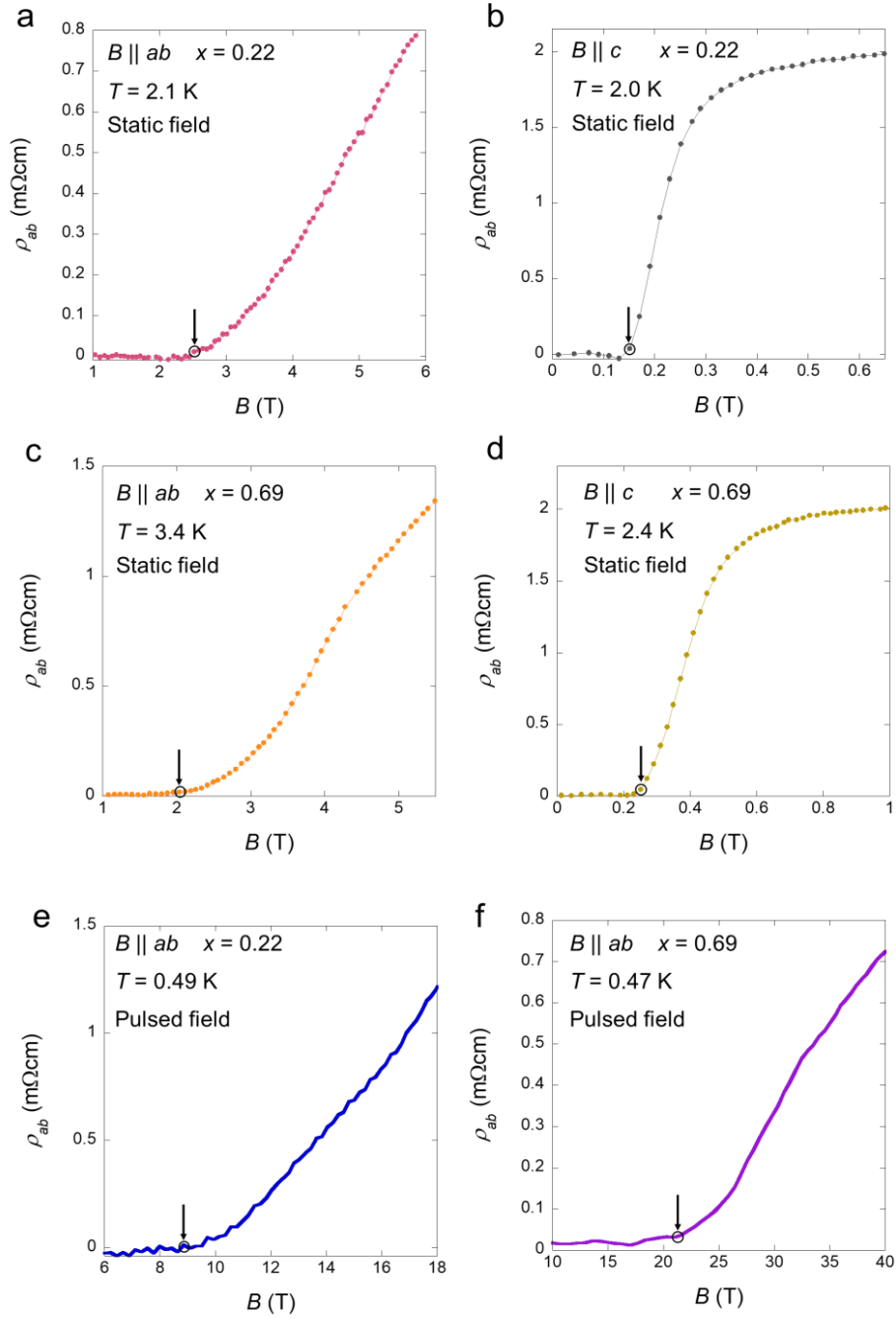

Supplementary Fig. 3 Examples of the criteria for the beginning of the resistive increase from zero resistivity of the $\rho_{ab}(B)$ curves. **a–f** Field dependence of the resistivity by static field for $x = 0.22$ in the fields parallel to the *ab*-plane (**a**) and *c*-axis (**b**), for $x = 0.69$ in the fields parallel to the *ab*-plane (**c**) and *c*-axis (**d**), and by pulsed field for $x = 0.22$ (**e**) and $x = 0.69$ (**f**). The black arrows and open circles exhibit the beginning of the resistive increase from zero resistivity of the $\rho_{ab}(B)$ curves to estimate the upper critical fields for Supplementary Fig. 4.

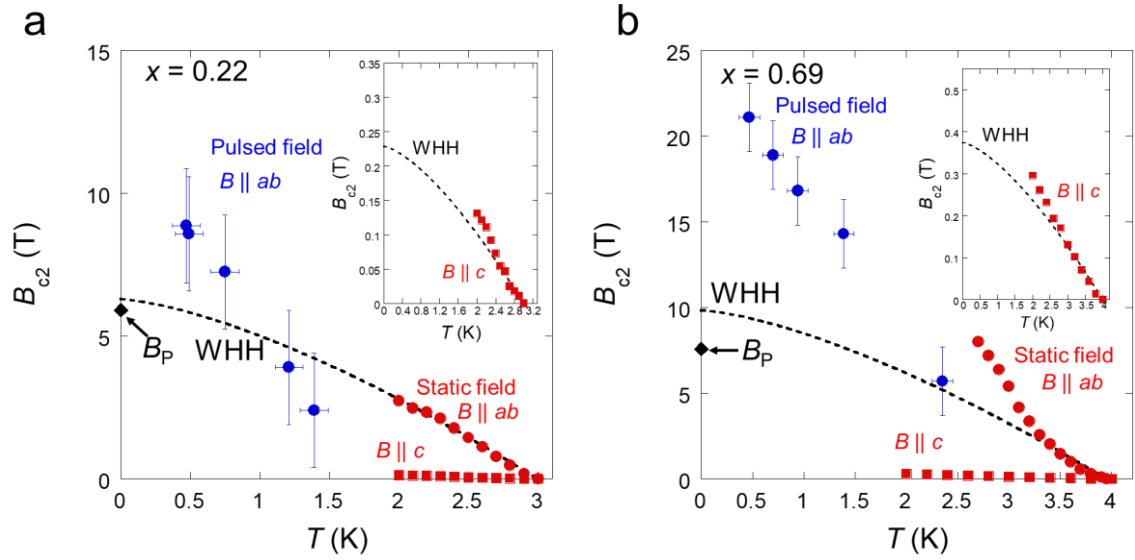

Supplementary Fig. 4 $B_{c2}$ determined from the beginning of the resistive increase from the zero resistivity of the $\rho_{ab}(B)$ curves. **a, b** Temperature dependence of the upper critical fields estimated from the beginnings of the resistive increase from zero resistivity of the $\rho_{ab}(B)$ curves for $x = 0.22$ (**a**) and $x = 0.69$ (**b**).

## S.3 Comparison with $B_{c2}$ estimated from $\rho_{ab}(T)$ and $\rho_{ab}(B)$

In order to compare with the upper critical fields determined from $\rho_{ab}(T)$ and $\rho_{ab}(B)$ curves, we show various upper critical fields in Supplementary Fig. 5. The upper critical fields estimated from $\rho_{ab}(T)$ curves are almost consistent with $\rho_{ab}(B)$ curves.

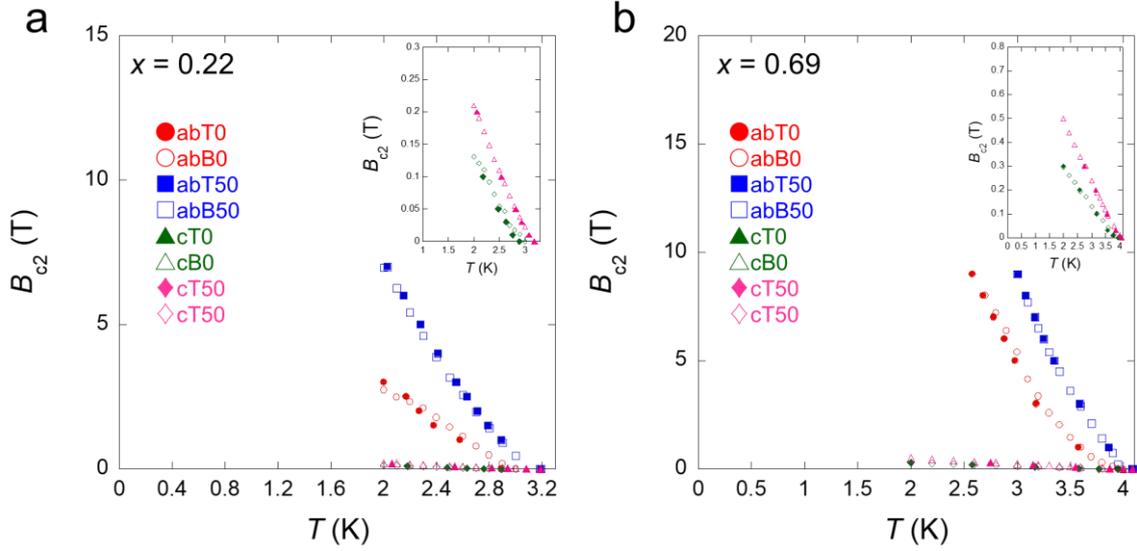

Supplementary Fig. 5 Comparison of $B_{c2}$ estimated from $\rho_{ab}(T)$ and $\rho_{ab}(B)$ curves. **a, b** Temperature dependence of the upper critical fields from the $\rho_{ab}(T)$ and $\rho_{ab}(B)$ curves for $x = 0.22$ (**a**) and $x = 0.69$ (**b**). The abT0 (red closed circles) shows the in-plane upper critical fields defined as the beginning of the resistive increase from zero resistivity of the $\rho_{ab}(T)$ data, the abB0 (red open circles) of the $\rho_{ab}(B)$ data, abT50 (blue closed squares) as the midpoint of the resistive transition of the $\rho_{ab}(T)$ and abB50 (blue open squares) of the $\rho_{ab}(B)$, respectively. The cT0, cB0, cT50, and cT50 show the out-of-plane upper critical fields from same way as the in-plane upper critical fields.